\documentclass{iauc}
\usepackage{graphicx}

\title[The Stellar Populations of dSph Galaxies in Nearby Groups] 
{The Stellar Populations of dSph Galaxies in Nearby Groups}

\author[Da Costa]   
{G. S. Da Costa}%

\affiliation{Research School of Astronomy \& Astrophysics, The Australian
National University \break 
email: gdc@mso.anu.edu.au}

\pubyear{2005}
\volume{198} 
\pagerange{1-6}
\date{\today and in revised form ??}
\jname{Near Field Cosmology with dwarf Elliptical Galaxies}
\editors{H. Jerjen and B. Binggeli, eds.}
\begin{document}

\maketitle

\begin{abstract}
In this contribution initial results are presented from a program 
to study in detail the stellar populations of dwarf Spheriodal (dSph) galaxies 
in three nearby groups.  The long-term aim of the program is to assess
the influence of environment in governing the evolution of these low-luminosity
systems.  Specific results described here include the detection and 
measurement of 
intermediate-age upper-AGB populations in dSphs in the M81 and Cen~A groups,
and the discovery that four of the five low-luminosity early-type dwarfs in the
low density Sculptor group contain modest amounts of neutral hydrogen gas.


\end{abstract}

\firstsection 
\section{Introduction}

We have known for more than two decades that
the dwarf Spheroidal (dSph) companion galaxies to the Milky Way show
a variety of star formation histories.  These
range from essentially a single old stellar population 
(e.g.\ Ursa Minor, \cite[Olszewski \& Aaronson 1985, Carrera et al.\ 2002]
{OA85,CA02}) through to the complex star formation histories of dSphs 
such as Carina 
(e.g.\ \cite[Monelli et al.\ 2003]{MM03}), Fornax 
(e.g.\ \cite[Saviane et al.\ 2000]{SH00}) and Leo~I 
(e.g.\ \cite[Gallart et al.\ 1999]{GF99}), which contain stars as young 
as $\sim$1 Gyr, or perhaps even younger.  The derivation of these star
formation histories is now based on
deep colour-magnitude (c-m) diagrams that reach below the main sequence turnoff
for the oldest populations, but it is important to recall that the first
evidence for the existence of intermediate-age (i.e.\ age 
between $\sim$1 and 10 Gyr) populations in dSphs came from the observation 
that some systems contain upper-AGB stars (cf.\ \cite[Aaronson \& 
Mould 1980]{AM80}).  Upper-AGB stars are stars 
with sufficient mass that they evolve to luminosities 
well above the tip of the red giant branch in the c-m
diagram.  Provided the underlying 
population is relatively metal-poor ($<$[Fe/H]$>$ $\leq$ --1.0, approximately),
the presence of upper-AGB stars, which are frequently seen as carbon stars, 
in a stellar population is an unambiguous signature of the existence of an 
intermediate-age population, and the luminosity of the brightest upper-AGB 
stars is a measure of the age of the youngest intermediate-age population 
present.   

With advent of HST it has also become possible to determine c-m diagrams that 
reach below the level of the horizontal branch for the 
dSph\footnote{Throughout this contribution the term `dSph' is synomous with
`low luminosity dE'.} 
companions to M31 (e.g.\ \cite[Da Costa et al. 2002]{GD02}), and thus gain
some first-order information on their star formation histories.  All the
systems studied contain RR Lyrae variables, indicating that they 
contain at least some stars with ages comparable to that of the Milky Way 
globular clusters (this is also true of the Milky Way dSph companions).
Nevertheless, the ubiquitous red horizontal branches in the c-m diagrams
 argue
for extended epochs of star formation and ``mean ages'' younger than that
of the Galactic globular clusters (much like the Galactic dSph Leo II -- see
\cite[Mighell \& Rich 1996]{MR96}).  However, the M31 dSph companions
appear to lack the substantial intermediate-age upper-AGB populations 
seen in many of the Milky Way's dSph companions.  Confirmed populations
of upper-AGB stars are known only in some of the M31 dSph companions, such
as And~II and And~VII (Cas) (see \cite[Harbeck et al.\ 2004]{DH04} and 
Harbeck, these proceedings), but the
luminosities of these stars suggest ages of perhaps 6--9 Gyr, in contrast
to the younger (and more luminous) stars seen in Carina, Leo~I and Fornax
(cf.\ \cite[Da Costa et al. 2000]{GD00}).

Within the Local Group there are a further four known low-luminosity early-type
dwarf galaxies.  These are the transition-type (dIrr/dE) systems Phoenix
and LGS~3, that are outlying companions of the Milky Way and M31, 
respectively, and the relatively isolated dSph systems Tucana and Cetus.
Both Phoenix and LGS~3 have modest populations of young stars, contain
small amounts of H{\small I} gas, and clearly have had star
formation from the distant past to the present (cf.\ \cite[Holtzman et al.\
2000, Miller et al.\ 2001]{HS00,BM01}).  The somewhat limited data for
Cetus shows that this dSph has a red horizontal branch morphology 
(\cite[Sarajedini et al.\ 2002]{AS02}), and that it lacks any obvious
upper-AGB population (\cite[Harbek et al.\ 2004]{DH04}).  It is apparently
broadly
analogous to the M31 dSph companions.  The isolated dSph Tucana, however,
shows a relatively strong blue horizontal branch morphology in
its c-m diagram (cf.\ \cite[Da~Costa 1998]{GD98}); indeed other than the 
Galactic dSph companion 
Ursa Minor, the horizontal branch morphology of Tucana is the bluest known
of the Local Group dSphs, suggesting a relatively brief star formation 
epoch in the distant past.  A recent determination of the radial velocity
of Tucana (Tolstoy, personal communication) has removed any possibility
of an association between this dSph and an H{\small I} cloud that lies nearby
on the sky (cf.\ \cite[Osterloo et al.\ 1996]{OD96}).  Given this result, the 
\cite{OD96} upper limit of M$_{HI}$/L$_{B}$ $\approx$ 10$^{-2}$ in solar 
units indicates that the Tucana dSph is similar to the Galactic dSphs in
lacking neutral gas.

Do we understand what causes this variety of star formation histories in
early-type low-luminosity galaxies?  In general the answer is `no' but it
is clear from the existing results that `environment', represented by, for
example, proximity to a luminous galaxy, and the type of that galaxy, plays
a role in influencing the star formation histories of these 
systems.  \cite{vdB94} noted that for the Galaxy's dSph companions, there
is a tendency for the systems with larger intermediate-age component to
lie at larger galactocentric distances.  The same trend may also apply for 
the M31 dSphs, as the two systems with likely intermediate-age populations,
And~II and And~VII (Cas), are among the most distant from M31.  Further, the
difference in relative frequency of intermediate-age populations between
the Galactic and M31 dSph systems may have its origin in the different
`parent' galaxy types: M31 has a much larger bulge than the Galaxy.

Clearly if we wish to understand the role of environment in influencing
the evolution of these low luminosity systems, we need to study the 
stellar populations of
dSphs in groups beyond the Local Group.  Fortunately, the nearest groups
provide a variety of different environments.  The
Sculptor group is a loose aggregation of galaxies ranging in distance
from $\sim$1.5 to $\sim$4 Mpc.  It contains half a dozen or so early-type
low-luminosity systems though the group contains mainly late-type galaxies.
In contrast, the M81 and Cen~A groups are relatively dense, compact
groups that lie at distances of $\sim$3.5 to $\sim$4~Mpc.  Both groups
contain at least a dozen or more low-luminosity early-type galaxies.  In
all three groups the dSphs cover a range of internal (absolute magnitude,
size, etc) and external (distance from nearest large galaxy, local galaxy
density) properties.

Together with my collaborators I am involved in a number of programs 
aimed at studying in detail the stellar populations of the dSphs in these
groups.  We are using HST ACS/WFC and existing `snapshot'
data to determine distances and metal abundances, Gemini-North NIRI and
VLT ISAAC near-infrared images to study upper-AGB populations, and, for
the two southern groups, Australia Telescope Compact Array and Parkes
radiotelescope observations to determine, or place limits on, neutral gas
contents.
In this contribution I present initial results for some of the dSphs in these
groups.

\section{M81 Group}

The first investigation of the stellar populations of dSph galaxies in
this group was that of \cite{NC98}, who used multi-orbit HST/WFPC2 $V$ and $I$
imaging to determine c-m diagrams for two M81 group dSphs, F8D1 and BK~5N\@.
The latter system has M$_{V}$ $\approx$ --11.3 and lies 
approximately 70kpc from M81 and $\sim$30kpc from NGC~3077 (in projection).
The \cite{NC98} c-m diagram clearly shows that an upper-AGB population is 
present in this galaxy, indicating on-going star formation.  
\cite{NC98} estimate the upper-AGB termination luminosity as M$_{bol}$
$\approx$ --4.3, which suggests that stars formed in this dSph until 
$\sim$8 Gyr ago.  F8D1, on the other hand, is a luminous 
(M$_{V}$ $\approx$ --14.2)
low surface brightness dSph that lies $\sim$120kpc from M81 and approximately
35kpc from the spiral galaxy NGC~2976 (in projection).  The c-m diagram
shows a strong upper-AGB population, which remains even when
allowance is made for the likely presence of 47~Tuc-like long period variables,
i.e.\ stars that lie above the red giant branch tip, but which are not of
intermediate-age.  Such stars can occur whenever the mean metallicity of
a system exceeds [Fe/H] $\approx$ --1.0 dex.  \cite{NC98} indicate that the 
most luminous of the upper-AGB stars in F8D1 have M$_{bol}$ $\approx$ --5.0.
This dSph has also been observed in the near-infrared using
NIRI on the Gemini-N telescope.  The observations (see \cite[Da~Costa 2004
for a full discussion]{GD04}) yield very similar results to those of
\cite{NC98}: the most
luminous stars again have M$_{bol}$ $\approx$ --5.0 and the
luminosity functions for the upper-AGB stars calculated from the I-band
and K-band data agree well (\cite[Da~Costa 2004]{GD04}).  Comparisons with
LMC and SMC cluster data, and with near-IR data for the Galactic dSph
companion Leo~I (cf. \cite[Menzies et al. 2002]{MW02}), suggest that F8D1 
contains stars as young as 1--2 Gyr.

Gemini-N NIRI imaging has now also been analysed for an additional M81 group 
dSph, known as kkh57 (Da~Costa et al., in preparation).  This dSph is a small, 
low luminosity 
(M$_{V}$ $\approx$ --10.9) system, that lies in a rather isolated location 
in the group, approximately 370kpc in projection from M81\@.  The c-m diagram
for kkh57 clearly shows the presence of upper-AGB stars,
indicating on-going star formation in this dSph as well.  The brightest
of the upper-AGB stars have M$_{bol}$ $\approx$ --4.3 to --4.6, somewhat 
more luminous than for BK~5N yet fainter than for F8D1\@.  This would appear
to be prima facie evidence that the diversity of star formation histories
seen among the Local Group dSphs is also present in the dSphs of the 
M81 group.

\section{Cen A Group}

\begin{figure}
\includegraphics[width=8cm,angle=-90]{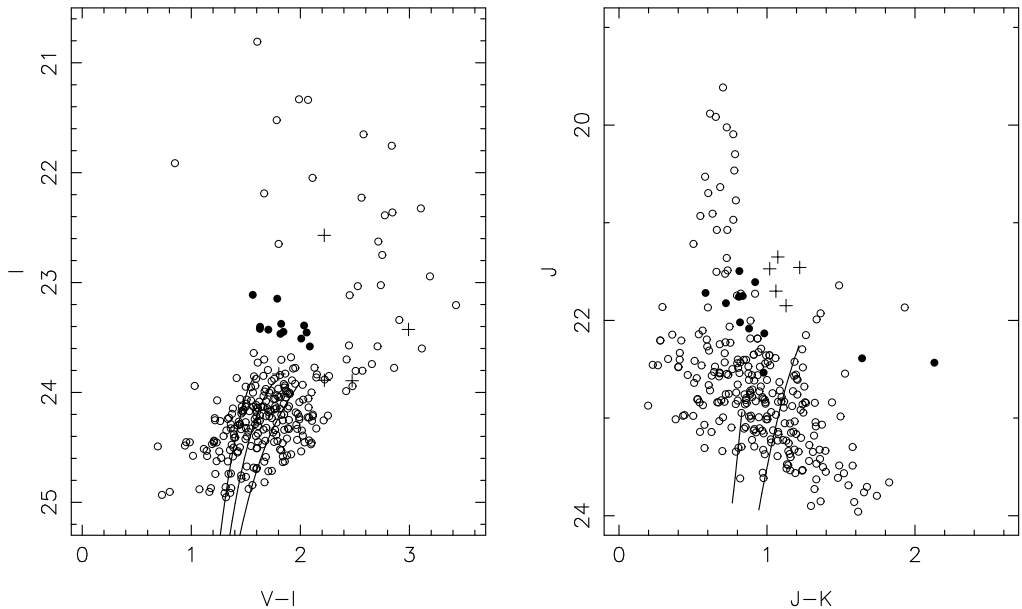}
  \caption{Colour-magnitude diagrams for the Cen~A group dSph AM1343-452
(kk217).  The left panel shows $V,I$ photometry derived from HST snapshot 
images. Shown also are the red giant branches (RGBs) of the Galactic globular
clusters M15 ([Fe/H]=--2.12), M2 (--1.62) and NGC~1851 (--1.22) shifted to
the distance modulus and reddening of the dSph.   The right panel shows 
$J, K$ photometry derived from VLT ISAAC images together with the RGBs for
the globular clusters M92 (--2.28) and 47~Tuc (--0.76).  In both panels
the stars plotted are those measured on both sets of data.  A
number of potential upper-AGB stars are plotted as filled symbols, for
those selected in the left panel, and as plus-signs, for those selected
in the right panel.  The same symbol is used to show the location of these
stars in the corresponding diagram.} \label{CenAfig}
\end{figure}

In this group we have recently obtained VLT ISAAC near-IR imaging of
14 dwarf systems (see Rekjuba et al., these proceedings).  Analysis of this 
dataset
continues but I present here initial results for AM1343-452 (kk217), a dSph
with absolute visual magnitude M$_{V}$ $\approx$ --11.5 that lies 
approximately 290 kpc in projection from Cen~A\@.  \cite{Ka02} classify
it as a Cen~A companion.  The left panel of Fig.\ \ref{CenAfig} shows an
optical c-m diagram derived from the HST snapshot data for this
galaxy.  {\it The stars plotted are only those that are measured on both
the HST and the VLT ISAAC data.}   
The tip of the red giant branch is clearly visible at $I$ $\approx$
23.9 $\pm$ 0.1.  Overplotted are standard globular cluster giant branches 
which 
suggest a mean metallicity for the dSph of $<$[Fe/H]$>$ $\approx$ --1.55
$\pm$ 0.3 dex.  The field contamination is substantial, given the comparatively
low galactic latitude, but nevertheless there is some indication of a 
population of stars that lie above the red giant branch tip.  A sample of such
potential upper-AGB stars, selected using the c-m diagram of BK~5N of 
\cite{NC98} as a guide, are plotted as filled symbols in this panel.

The right panel of Fig.\ \ref{CenAfig} shows 
the near-IR colour magnitude 
diagram derived from the ISAAC data. Plotted in this panel are the red giant 
branches of the Galactic globular clusters M92 and 47 Tuc, using the same 
AM1343-452 reddening and distance modulus as for the left panel.  In this 
c-m diagram most of the 
Galactic field stars have colours of $J-K \approx 0.7$, and so form a 
vertical sequence in the c-m diagram.  Five possible upper-AGB stars 
are plotted as plus symbols in this panel: the left panel shows that these
stars also fall at or above the RGB tip in the optical c-m diagram.  Similarly,
while some of the upper-AGB candidates selected in the left panel fall
in the field star sequence in the near-IR c-m diagram, the majority
are above the RGB tip and two have quite red near-IR colours.  These two
stars may well be carbon stars.  
 
Regardless of the actual status of individual stars, these data suggest
that AM1343-452 does contain a population of upper-AGB stars.  This 
interpretation
confirmed by an analysis of the field-subtracted J-band luminosity function
(see Rejkuba et al., these proceedings), which reveals a clear excess of 
stars brighter than  the RGB tip.  The most luminous of these
stars have M$_{bol}$ values in the range of --4.5 to --4.8, corresponding to
ages of a few Gyr.  Clearly there has been on-going star formation 
in this particular Cen~A group dSph.

While the complete data set remains to be analysed in detail, it appears
that a sizeable fraction (perhaps as much as 50\%) of the Cen~A group 
dwarfs observed in the near-IR show evidence for upper-AGB populations, and
thus on-going star formation.  Nevertheless, there
are clearly some dSphs in our sample that apparently lack such populations.  
Further modelling of the image
completeness as a function of magnitude and colour is required to be 
certain of this result, but it is likely that some of the
dSphs in this group lack intermediate-age populations.  These would be
dSphs in which essentially all the stars are older than 8--10 Gyr, making
them Cen~A group analogues to Local Group systems such as Sculptor and
And~I\@.  When the full analysis is complete it will be interesting to see 
where these particular dwarfs lie in the group, as compared to the locations
of the dSphs with intermediate-age populations.
 
\section{Sculptor Group}

In this group there are five low-luminosity early-type galaxies with confirmed
group membership.  Four of these five have been classified by \cite{JB00} on
the basis of their morphology: three are classed as transition types
while the fourth is classed as a dwarf elliptical.  The fifth system is
classified as a dSph by \cite{Ka00}.  
We do not as yet have any near-IR imaging data for these systems, though HST 
snapshot imaging is available for all five.  These data, together with 
existing ground-based optical imaging data for ESO~540--032,
reveal the presence of blue stars in the central regions of at least two of 
the dwarfs (cf.\ \cite[Karachentsev et al.\ 2000, Jerjen \& Rejkuba 
2001]{Ka00,JR01}), suggesting that they have had modest levels of 
relatively recent star formation.

We have observed all five systems with the Parkes radio telescope
to determine, or place limits on, their neutral hydrogen contents.  The
results are described in more detail in Bouchard et al.\ (these preceedings)
but in summary, four of the five systems, including all three classified
as transition types, and including both of the systems with central blue stars,
were detected in H{\small I}\@.  The M$_{HI}$/L$_{B}$ values range
from 0.08 to 0.19 in solar units for the detections, comparable to
those for the Local Group transition systems LGS~3 and Phoenix, while the 
3$\sigma$ upper limit for the non-detection corresponds to 
M$_{HI}$/L$_{B}$ $<$ 0.08 in solar units.  For three of these four systems,
knowledge of the H{\small I} radial velocity
has allowed the generation of H{\small I} maps from data available in the
Australia Telescope Compact Array archive.  These maps show that for
two systems, the center of the H{\small I} distribution is offset from 
that of the optical image, while in the third, the H{\small I}, which is 
apparently quite extended, peaks near the center of the optical image
(Bouchard et al., in preparation).  Again these situations are reminiscent of
that for the Local Group transition objects Phoenix and LGS~3, but considerably
more information than is currently available is required before we can attempt
to connect the gas dynamics with the recent star formation histories (cf.\
\cite[Gallart et al. 2001]{CG01}).

\section{Conclusions}  
  
Although the sample of objects analysed so far is not large, it does seem
clear that in the M81 and Cen~A groups a substantial fraction of the dSphs
have had on-going star formation, and that like the Local Group, there is
considerable diversity in the extent and duration of this phenomenon.
On the other hand, in the Scl group, which is a less dense environment,
the low-luminosity early-type galaxies are predominantly of the transition
type, possessing modest amounts of neutral gas and showing indications of
relatively recent star formation.  This would appear to be a direct 
observational confirmation of the premise of \cite{GG03}, who suggested
that ``transition-type dwarfs $\dots$ should replace dSphs in
isolated locations where stripping is ineffective''.

In a more general sense these results support the idea that `environmental
conditions' do play an important role in governing the evolution of 
low-luminosity early-type systems.  Specificially, the majority of the
Scl group objects have apparently
evolved in a relatively independent fashion, generating a star formation
history that extends to the present epoch. The M81 and Cen~A group
objects, on the other hand, have apparently been influenced to a 
greater or lesser extent by external
factors depending on their orbits and location within the group.  It will
be interesting to see the extent to which these conclusions remain valid 
once the full sample of dSph observations is analysed.   

\begin{acknowledgments}
GDaC would like to acknowledge the contributions of his collaborators,
particularly Marina Rejuka, in developing the results presented here.
GDaC's work on dwarf galaxies is supported in part by ARC Discovery Grant
DP0343156. 
\end{acknowledgments}

\newpage

\begin{discussion}

\discuss{Karachenstev}{There are two peripheric companions to M31: the
Cass dSph and the Cetus dSph.  Why do you use for Cetus its original name 
(given by its discovers), but in the case of Cassiopeia you ignore the
original name?}

\discuss{Da Costa}{Cetus is relatively isolated Local Group dSph, and as
far as I am aware, is not associated with M31.  There is therefore no reason
to refer to it by any other designation.  However, for the confirmed dSph
companions of M31, I prefer to follow the convention initiated by Sidney 
van den Bergh of referring to these systems as And $\sharp\sharp$, where 
$\sharp\sharp$ is a roman 
numeral.  It is important, however, to also include other designations where
these exist:
specifically And VI is also referred to as 
the Pegasus dSph (as distinct from the Pegasus dwarf irregular DDO~216) and
And~VII, discovered by Karachentsev \& Karachentseva (1999, A\&A, 341, 355)
is also referred to as the Cas dSph.} 

\discuss{Gallagher}{Are there any obvious trends in the overall mix of dwarf
galaxy types between the M81 and Cen~A groups, where recent interactions
are important, as compared with the Local and Sculptor groups which are
quiescent?}

\discuss{Da Costa}{There does exist, of course, the well established
morphology-density relation in which early-type dwarfs are more commonly
found in higher density environments than late-type dwarfs. However, 
comparing specific groups in this context is somewhat difficult because of 
completeness concerns.  Fig.\ 1 of Karachentsev
el al.\ (2002, A\&A, 383, 125) shows there are approximately equal
numbers of `Sph' and `Irr' dwarfs within the M81 group, with the `Sph'
galaxies on average somewhat closer to M81 than the `Irr' dwarfs.  
To a comparable magnitude limit, the Local Group also has roughly
equal numbers of early- and late-type dwarfs, with the early-type
dwarfs notably clustered around the Milky Way and M31.  Similarly,
judging from Table 2 of Karachentsev et al.\ (2002, A\&A, 385, 21) or
Fig.\ 2 of Jerjen et al.\ (2000, AJ, 119, 593), in the Cen~A group there
are about one-third more late-type dwarfs than early-types, though again
the early-type dwarfs are generally closer to Cen~A\@.  The Sculptor group
though is clearly different in this respect as it has roughly three times 
as many late-type dwarfs as early-types, as judged from Fig.\ 2 of 
Jerjen et al.\ (2000).
Our program, however, is endeavouring to go beyond these morphological
relations in attempting to assess more quantitatively possible relations 
between star formation histories and environment. }

\discuss{Caldwell}{You find that a larger number of dE and transition
dwarfs in Sculptor have been unperturbed over time when compared to those
in the M81 and Cen~A groups.  Given that there are more early-type dwarfs
in the latter two groups, could the observed difference in Sculptor be
a small number statistics problem?}

\discuss{Da Costa}{It is true that the analysis so far is based on 
relatively small numbers of objects, both in terms of the number of 
objects analysed in detail in the M81 and Cen~A groups, and in terms 
of the fact that there are only 5 low-luminosity early-type systems in the
Sculptor group.  Nevertheless I would be surprised if, for example, our
H{\small I} observations of the dE and dE/Irr systems in the Cen~A group
produced the same high relative fraction of detections as was found in the 
Sculptor group.} 
\end{discussion}

\end{document}